\documentclass[12pt]{article}

\font\grande=cmr9.5 scaled \magstep4
\font\medio=cmr9.5 scaled \magstep2
\outer\def\beginsection#1\par{\medbreak\bigskip
      \message{#1}\leftline{\bf#1}\nobreak\medskip
\vskip-\parskip
      \noindent}
\begin{document}
\bibliographystyle {unsrt}

\titlepage

\begin{flushright}
CERN-PH-TH/2014-063
\end{flushright}

\vspace{15mm}
\begin{center}
{\grande Faraday scaling and the Bicep2 observations}\\
\vspace{15mm}
 Massimo Giovannini 
 \footnote{Electronic address: massimo.giovannini@cern.ch} \\
\vspace{0.5cm}
{{\sl Department of Physics, Theory Division, CERN, 1211 Geneva 23, Switzerland }}\\
\vspace{1cm}
{{\sl INFN, Section of Milan-Bicocca, 20126 Milan, Italy}}
\vspace*{2cm}

\end{center}

\vskip 1.5cm
\centerline{\medio  Abstract}
As repeatedly speculated in the past, the linear polarization of the Cosmic Microwave Background can be rotated
via the Faraday effect. An economic explanation of the recent Bicep2 observations, not relying on long-wavelength tensor modes of the 
geometry, would stipulate that the detected B mode comes exclusively from a Faraday rotated E mode polarization.
We show hereunder that this interpretation is ruled out by the existing upper limits on the B mode polarization obtained by independent 
experiments at observational frequencies much lower than the operating frequency of the Bicep2 experiment. 
We then derive the fraction of the observed B mode polarization ascribable
to the Faraday effect and suggest a dedicated experimental strategy for its detection.
\noindent

\vspace{5mm}

\vfill
\newpage
The Bicep2 collaboration  \cite{BICEP2} recently measured  the B mode polarization for angular scales of the order of the degree and at a pivot frequency $\nu_{p}=150$ GHz:
\begin{equation}
{\mathcal G}_{B\,\ell} = \frac{\ell (\ell +1) }{ 2\pi } \,C_{\ell}^{(BB)} = {\mathcal O}(10^{-2})\,\,\mu\mathrm{K}^2,
\label{int}
\end{equation}
where ${\mathcal G}_{\ell B}$ denotes the angular power spectrum of the B mode autocorrelation, $\ell\simeq \pi/\vartheta$ is the multipole moment and $\vartheta$ is the angular scale. 
If the whole effect is attributed to the primordial tensor modes, amplified by the pumping action of the gravitational field in a (spatially flat) Friedmann-Robertson-Walker geometry, the present frequency 
of the corresponding gravitons would be ${\mathcal O}(10^{-17})$ Hz.

In what follows the sole assumptions will be instead that the the tensor modes of the geometry are absent and that whole Bicep2 signal comes from the Faraday rotation of the polarization plane of the Cosmic Microwave Background 
(CMB in what follows).  If this is the case it will be shown that we are led into an interesting contradiction: direct upper limits 
on the B mode polarization at frequencies $\nu \ll \nu_{p}$ are violated if the detected signal is ascribed to a purported Faraday effect.
The  B mode polarization induced by the Faraday rotation scales with the frequency. Conversely the B mode induced by the tensor modes of the geometry 
is frequency independent.

In the recent past, direct upper limits on the B mode autocorrelation ${\mathcal G}_{\ell B}$ have been presented, over different observational frequencies, by various collaborations aiming at a direct detection of the CMB polarization. The pivot frequencies of the CMB polarization experiments can be conventionally divided into two ranges conventionally denoted hereunder by $\nu_{low}$ and $\nu_{high}$:
\begin{equation}
26 \,\, \mathrm{GHz} \leq \nu_{low} \leq 36\,\, \mathrm{GHz}, \qquad
100 \,\, \mathrm{GHz} \leq \nu_{high} \leq 150\,\, \mathrm{GHz}.
\label{nulowhigh}
\end{equation}
The Dasi (Degree Angular Scale Interferometer) \cite{dasi} and the Cbi (Cosmic Background Imager) \cite{cbi} experiments 
were both working in a range coinciding exactly with $\nu_{low}$. 
Four other experiments have been conducted around $\nu_{high}$ and they are:
{\it i)} Boomerang (Balloon Observations of Millimetric Extragalactic Radiation and Geophysics) working at $145$ GHz with 
four pairs of polarization sensitive bolometers \cite{boom}; {\it ii)}
Maxipol (Millimiter Anisotropy experiment Imaging array) working at $140$ GHz with 12 polarimeters
 \cite{maxipol}; {\it iii)} Quad\footnote{An acronym or a contraction between the Quest (Q and U extragalactic sub-mm telescope) and the Dasi experiments.}
working with 31 pairs of polarization sensitive bolometers: 12 at $100$ GHz and $19$ at $150$ GHz \cite{quad}; 
{\it iv)} Bicep2 \cite{BICEP2} and its precursor Bicep1 \cite{bicep1}  work, respectively, at $150$ GHz and $100$ GHz.

There are finally three polarization sensitive experiments working in mixed or intermediate frequency ranges. They include:
{\it a)} the WMAP experiment \cite{WMAP9} (Wilkinson Microwave Anisotropy Probe) \cite{WMAP9} spanning five frequencies from $23$ 
to $94$ GHz; {\it b)} the Capmap experiment (Cosmic Anisotropy Polarization Mapper) \cite{capmap}, with $12$ receivers operating between $84$ and $100$
GHz and four receivers operating between $35$ and $46$ GHz; {\it c)} the Quiet (Q/U imager experiment) \cite{quiet} operating at $43$ GHz (during the first season of the experiment) and at $95$ GHz (during the second season of the experiment).

The dispersion relations governing the propagation of electromagnetic excitations in a cold, electrically neutral and magnetized plasma \cite{plasma}
determine the angular shift  $\Delta \varphi$ of the polarization plane of the linearly polarized wave propagating, for instance, along the 
third Cartesian axis:
\begin{equation}
\Delta\varphi = \frac{\omega_{Be}}{2} \biggl(\frac{\omega_{pe}}{\omega}\biggr)^2 \Delta z,
\label{int2}
\end{equation}
showing that the rotation rate (i.e. $\Delta \varphi/\Delta z$) decreases as the inverse square of the angular frequency of the photons (i.e. $\omega = 2 \pi \nu$)
and it increases linearly with the magnetic field intensity, i.e. with the Larmor frequency of the electrons.
The result (\ref{int2}) is approximate and it holds as long as 
the plasma and Larmor frequencies of the electrons (i.e. $\omega_{pe}$ and $\omega_{Be}$) are both smaller than 
the angular frequency of the CMB photons\footnote{During the formation 
of the CMB polarization  $\omega_{Be} = 1.7 \times 10^{-2}\,\, \mathrm{Hz}$ (for typical magnetic field strengths ${\mathcal O}(\mathrm{nG})$, $\omega_{pe} = 0.285\,\, \mathrm{MHz}$ for the fiducial set 
of WMAP9 parameters alone\cite{WMAP9}). For frequencies of the photons larger than few GHz the hierarchy between the scales of the problem is $\omega \gg  \omega_{pe} \gg \omega_{Be}$. The plasma and Larmor frequencies of the ions do not play any role and are both smaller than the corresponding frequencies of the electrons.}.

In the concordance cosmological lore the linear polarization is provided by the scalar adiabatic mode whose existence and properties 
have been firstly scrutinized by the WMAP experiment during the past decade \cite{WMAP9} (see for instance \cite{intr} for an introduction to the polarization of the CMB). The WMAP experiment 
discovered the E mode polarization whose existence has been later confirmed by different experiments (see e.g. \cite{quad}).  
The rotation rate of Eq. (\ref{int2}) determines a rotation in the Stokes parameters implying that the induced B mode polarization 
can be computed in terms of the E mode polarization  generated by the scalar adiabatic mode and measured by WMAP experiment.  
The details of this process have been studied in the past (see e.g. the first eight papers of Ref. \cite{far2})
but are not essential here. The only crucial piece of information will be the Faraday scaling law stipulating that the B mode polarization 
induced by the Faraday effect scales as the inverse of the fourth power of the frequency (or, which is the same, as the fourth power of the wavelength). Given the signal at a certain pivot frequency $\nu_{p}$ (that we shall choose to be the Bicep2 frequency),
the B mode polarization induced by Faraday rotation at a different observational frequency $\nu$ can be obtained in terms of  
 this simple scaling law (see e.g. \cite{far1} and third, fourth and fifth papers of Ref.\cite{far2}):
 \begin{equation}
{\mathcal G}_{\ell B}(\nu) = \biggl(\frac{\nu_{p}}{\nu}\biggr)^4 {\mathcal G}_{\ell B}(\nu_{p}),
\label{eq1}
\end{equation}
where $\nu_{p} = 150\, \mathrm{GHz}$ is the pivot frequency that coincides with the Bicep2 frequency.
The characteristic frequency dependence of Eq. (\ref{eq1}) arises since the Faraday generated B mode is quadratic in the rotation rate.
According to Eq. (\ref{eq1}), for frequencies larger than $\nu_{p}$ the B mode is suppressed while for frequencies 
$\nu<\nu_{p}$ the angular power spectrum is enhanced. At a given frequency (be it for instance $\nu_{p}$) 
the B mode power spectrum induced by the Faraday 
rate first increases with the multipole $\ell$, it has a maximum for $\ell \sim 1000$ and then it decreases again \cite{far2}.
The maximum around $\ell = {\mathcal O}(10^{3})$ is actually connected with the maximum of the 
E mode autocorrelation occurring for $\ell \sim 1000$ \cite{WMAP9}. 

Let us therefore use our main assumption and posit that the entire signal observed by Bicep2 comes predominantly from a Faraday rotated E mode polarization. This assumption requires that ${\mathcal G}_{\ell B}(\nu_{p})$ appearing in Eq. (\ref{eq1}) can be 
replaced by the angular power spectrum observed by Bicep2 since the 
correction coming from gravitational lensing is not essential for the angular scales considered hereunder.
Consider then, for sake of concreteness, two specific ranges of multipoles scrutinized by the Bicep2 
collaboration, 
\begin{eqnarray}
{\mathcal G}_{\ell B}(\nu_{p}) = 3.37 \times 10^{-2}\,\mu\mathrm{K}^2, \qquad  196 \leq \ell \leq 230,
\label{obs1}\\
{\mathcal G}_{\ell B}(\nu_{p}) = 5.07 \times 10^{-2}\,\mu\mathrm{K}^2,  \qquad  231  \leq \ell \leq 265.
\label{obs2}
\end{eqnarray}
It turns out that the ranges of Eqs. (\ref{obs1}) and (\ref{obs2}) overlap with the observations of other experiments producing upper limits 
on the B mode polarization at lower frequencies compatible with $\nu_{low}$.  In particular 
the Dasi collaboration provided upper limits for the B mode polarization implying \cite{dasi}
\begin{eqnarray}
{\mathcal G}_{\ell B}(\nu_{low}) &<& 2.12 \,\mu\mathrm{K}^2, \qquad 28\leq \ell < 245,
\label{UL1}\\
{\mathcal G}_{\ell B}(\nu_{low}) &<& 6.45 \,\mu\mathrm{K}^2, \qquad 246\leq \ell < 420,
\label{UL2}
\end{eqnarray}
at $95\%$ confidence level. The Faraday scaling law of Eq. (\ref{eq1}) can now be used with the purpose of computing ${\mathcal G}_{\ell B}(\nu_{low})$ 
for the particular values of Eqs. (\ref{obs1})--(\ref{obs2}). The result of this simple rescaling shall then be compared them with Eqs. (\ref{UL1})--(\ref{UL2}). 
Therefore, using  Eqs. (\ref{obs1})--(\ref{obs2}) into Eq. (\ref{eq1}) the B mode power spectrum at $\nu_{low}$ becomes:
\begin{eqnarray}
{\mathcal G}_{\ell B}(\nu_{low}) = 3.37 \times 10^{-2} \biggl(\frac{\nu_{p}}{\nu_{low}}\biggl)^4 \,\mu\mathrm{K}^2, \qquad  196 \leq \ell \leq 230,
\label{obs1a}\\
{\mathcal G}_{\ell B}(\nu_{low}) = 5.07 \times 10^{-2}\,\biggl(\frac{\nu_{p}}{\nu_{low}}\biggl)^4 \,\mu\mathrm{K}^2,  \qquad  231  \leq \ell \leq 265.
\label{obs2a}
\end{eqnarray}
Since we are discussing the bounds on a signal that decreases as $\nu$ gets larger, 
 the most constraining frequencies compatible with the observational range  are the lowest ones. Consequently
if we estimate $\nu_{low} = 26 \,\mathrm{GHz}$,  Eqs. (\ref{obs1a}) and (\ref{obs2a}) imply, respectively, 
\begin{eqnarray}
{\mathcal G}_{\ell B}(\nu_{low}) = 37.33\,\mu\mathrm{K}^2, \qquad  196 \leq \ell \leq 230,
\label{obs1b}\\
{\mathcal G}_{\ell B}(\nu_{low}) =56.16\,\mu\mathrm{K}^2,  \qquad  231  \leq \ell \leq 265.
\label{obs2b}
\end{eqnarray}
Equations (\ref{obs1b}) and (\ref{obs2b}) violate the Dasi upper limits of  Eqs. (\ref{UL1})--(\ref{UL2}): instead of 
${\mathcal O}(2)$ and ${\mathcal O}(6)\,\,\mu\mathrm{K}^2$ we get, from Eqs. (\ref{obs1b}) and (\ref{obs2b}),
${\mathcal O}(40)$ and ${\mathcal O}(60)\,\,\mu\mathrm{K}^2$ in an overlapping range of multipoles. 

Moving towards smaller angular scales but always remaining around $\nu_{low}$, the constraints on a purported Faraday rotation signal
may become even stronger once the direct measurements of ${\mathcal G}_{\ell B}$ will be available for larger multipoles (i.e. $\ell \sim 10^{3}$). Indeed, over these angular scales,  Cbi \cite{cbi} obtained 
an upper limit at $95\%$ confidence level implying ${\mathcal G}_{\ell B}(\nu_{low}) < 3.76\,\,\mu\mathrm{K}^2$.
This bound may be even more constraining than the Dasi limits since it corresponds to larger $\ell$ where  
we do know, on a theoretical ground, that the potential Faraday rotation signal will certainly be larger, i.e. ${\mathcal G}_{\ell B} \gg {\mathcal O}(10^{-2})\,\, \mu\mathrm{K}^2$. Hence the obtainable limit is potentially stronger, but, at the moment observationally weaker 
since direct measurements of the B mode power spectrum are still lacking for these multipoles. 

The same strategy discussed for $\nu = {\mathcal O}(\nu_{low})$  can be used to analyze the upper limits reported by other experiments operating at 
$\nu = {\mathcal O}(\nu_{high})$ or 
even at intermediate frequencies $\nu_{low} < \nu < \nu_{high}$. As it can be easily argued thanks to the Faraday scaling law of Eq. (\ref{eq1}), 
the tensions between 
the upper limits and the Bicep2 measurement get progressively less pronounced as $\nu$ approaches $\nu_{p}$. For instance between Bicep2 and Bicep1
(located at $\nu \sim 100$ GHz) the gain in frequency from $\nu_{p}$ will be given by a comparatively smaller factor, i.e. $(150/100)^4 = 5.06$. 
As a representative of the experiments working in a mixed frequency range, it is useful to mention the WMAP bound stipulating  ${\mathcal G}_{\ell B}(\nu_{V})< 0.25 \, \mu\mathrm{K}^2$ for $50 \leq\ell \leq100$ and for 
$\nu_{V} = 41$ GHz (the so called V band) \cite{WMAP9}. 
In an overlapping range of multipoles the Bicep2 data would imply ${\mathcal G}_{\ell B}(\nu_{p}) = (1.33 \pm 0.17) \times 10^{-2} \mu\mathrm{K}^2$ 
for $56\leq\ell\leq 90$.  Again, using Eq. (\ref{eq1}) we get that the Bicep2 signal rescaled at frequency $\nu_{V}$ becomes $2.38  \, \mu\mathrm{K}^2$
so that the WMAP upper limit is violated.  The WMAP collaboration has also a stronger bound on ${\mathcal G}_{\ell B}$ for $2 \leq \ell \leq 7$
but it cannot be used for the present purposes due to lack of measurements by the Bicep2 collaboration in that specific range.

While this analysis was in progress some authors speculated that the observed B mode polarization is the result of a primordial 
Faraday rotation of the E mode polarization (see last paper of \cite{far2}). It is not clear how the authors produce their estimates but it is 
nonetheless claimed that some specific models can produce a signal compatible with the Bicep2 observations. A byproduct of 
the present analysis implies that, if these models exist at all, they will be in patent contradiction with the existing upper limits on the B mode polarization
at lower frequencies, as explained above in general terms.

All in all,  the Bicep2 measurements together with the other bounds on the polarization observables demand that the fraction 
of B mode polarization potentially coming from a purported Faraday rotation of the E mode can be, at most, ${\mathcal O}(10^{-4}) \,\mu\mathrm{K}^2$
for angular scales of the order of the degree. This result 
suggests the possibility of a mixing between the tensor B mode and the Faraday rotated E mode that will be independently investigated.
The derived bound can be sharpened either by considering smaller angular scales or by 
improving the observational bounds at low frequencies. 
The Planck satellite did not publish yet any polarization data and the available analyses borrow the polarization 
observables from the WMAP data. The Planck instrument is composed by a low frequency instrument (LFI) (with frequencies between $30$ and $70$ GHz) 
while the remaining six frequencies are located between $100$ and $857$ GHz and form the high frequency instrument (HFI). 
It is unclear from the published calibrations of the LFI and of the HFI \cite{planck} if the sensitivity to the  B mode polarization will be sufficient for a detection.
Barring for specific experimental considerations,  it will be extremely important if the three low frequency channels of the LFI  could obtain explicit bounds on the B mode polarization not only for angular scales of the order of the degree but also for sub-degree scales possibly improving the Dasi and Cbi bounds. By going at smaller angular scales (i.e. higher multipoles) 
the B mode signal is theoretically rather well understood both analytically and numerically. It will therefore be possible to infer rather robust bounds on the magnetic field strength independent on any other theoretical consideration. 

Summarizing the gist of the argument we can say that  if the B mode autocorrelation arises as a Faraday rotated E mode, then, the B mode angular 
power spectrum scales with the frequency. If we identify the measured B mode power spectrum at $\nu_{p}=150$ 
GHz with the Faraday rotated E mode, the corresponding signal will be larger for frequencies lower than $\nu_{p}$.
 By rescaling the result from $\nu_{p}$ to $\nu_{low}$ we are in condition of checking 
the existing upper limits on the B mode polarization at lower frequencies.
Since the observational upper limits are violated, the main (and sole) assumption of the analysis is fallacious and the observed B mode polarization 
cannot come predominantly from a rotated E mode polarization. In the light of the current bounds at different frequencies and different angular scales 
we can say that the fraction of B mode polarization potential attributable to a Faraday effect is at most ${\mathcal O}(10^{-4}) \,\mu\mathrm{K}^2$ for $\nu 
= {\mathcal O}(150)$ GHz and degree angular scales.  
Further analyses of the Faraday scaling of CMB polarization observables, in particular at low frequency, 
could open a new observational window for a closer scrutiny of the Faraday effect of the CMB and of large-scale magnetism prior to photon decoupling.
In this respect, specific observational strategies have been suggested and we hope they will be soon considered by 
the forthcoming polarization experiments.

\end{document}